\documentclass[11pt]{article}
\usepackage[margin=1in]{geometry}
\usepackage{amsmath}
\usepackage{graphicx}
\usepackage{multicol}
\usepackage[colorlinks=true,linkcolor=black,urlcolor=blue,citecolor=blue]{hyperref}
\usepackage{enumitem}
\usepackage{xcolor}
\usepackage{colortbl}
\usepackage{booktabs}
\usepackage[backend=biber, style=authoryear]{biblatex}
\usepackage{threeparttable}
\usepackage{tocloft}
\usepackage{float}
\usepackage{wasysym}
\usepackage{multirow}
\usepackage{array}
\newcolumntype{C}{>{\centering\arraybackslash}p{1cm}}
\newcommand{\fullcirc}{\CIRCLE}
\newcommand{\halfcirc}{\LEFTcircle}
\newcommand{\emptycirc}{\Circle}
\usepackage{tikz}
\newcommand{\circradius}{0.34em}
\newcommand{\threequartercirc}{\tikz[baseline=-0.5ex]{%
		\begin{scope}
			\clip (0,0) circle (\circradius);
			\fill[black] (-0.34em,-0.34em) rectangle (0.16em,0.34em);
		\end{scope}
		\draw (0,0) circle (\circradius);}}
\newcommand{\quartercirc}{\tikz[baseline=-0.5ex]{%
		\begin{scope}
			\clip (0,0) circle (\circradius);
			\fill[black] (-0.34em,-0.34em) rectangle (-0.16em,0.34em);
		\end{scope}
		\draw (0,0) circle (\circradius);}}
\newcommand{\na}{\textemdash}

\begin{document}
	
	\begin{center}
		{\LARGE\bfseries How to Catch a GPU: A Taxonomy of Verification and Enforcement Mechanisms for International AI Agreements}
		
		\vspace{0.4in}
		
		{\bfseries Raymond Koopmanschap \quad Otto Barten}\\
		Existential Risk Observatory\\
	\end{center}

	\begin{abstract}
		Several international agreements have been proposed to regulate frontier AI development in response to catastrophic risks. However, there is no structured way to evaluate whether these proposals are enforceable, to assess where they might fail in practice, or to determine which combination of policies is most effective.
		
		We propose a taxonomy based on the principle that wherever sufficient capacity exists to violate an agreement, it must be under a control regime. This decomposes the problem of ensuring compliance with the agreement into preventing uncontrolled resource acquisition, detecting all capacity outside the control regime, and preventing escape from the control regime. Existing proposals consist of individual policies that address one or more of these sub-problems. Because the compute required for dangerous capabilities may decrease over time, more actors can violate an agreement and enforcement of these policies becomes harder. We define the enforcement breaking point as the FLOP-threshold or equivalent metric at which a policy loses its effectiveness in solving the sub-problem, and introduce a set of factors to assess how and why this breakdown occurs. This reveals which of the three sub-problems any given proposal adequately addresses, and where enforceability breaks down first.
		
		Applying this taxonomy to existing proposals reveals that more focus is placed on preventing escape from the control regime, while preventing resource acquisition and detecting all capacity outside the control regime receive less attention. By making these gaps explicit, this taxonomy can help researchers and policymakers prioritize future enforcement and verification efforts.
	\end{abstract}
	
	\newpage
	\tableofcontents
	\newpage
	
	\section{Introduction}
	International AI agreements aimed at preventing catastrophic risks from frontier AI development have been proposed \parencite{scher2025international, aguirre2025keepfuture, alramiah2025globalregime, miotti2025narrow}. These agreements largely rely on controlling the hardware needed to build frontier AI systems. A growing literature proposes concrete measures for doing so \parencite{scher2025mechanisms, baker2025verifying, harack2025verification, wasil2024verification}, from chip tracking to on-site inspections.
	
	However, this literature does not assess where verification and enforcement become infeasible as the compute required for a violation shrinks. For example, while current detection methods can find a large data center, it is unclear whether they remain effective at a rack-sized facility.
	
	This paper proposes a taxonomy that decomposes verification and enforcement into three sub-problems: preventing escape from the control regime, preventing uncontrolled resource acquisition, and detecting hidden capacity. First, we introduce the taxonomy that describes how these sub-problems relate. Second, we map existing proposals from the literature onto them. Third, we analyze for each sub-problem whether its solution remains tractable as facility size decreases. From this analysis, we argue that detection of hidden capacity is the first to become intractable as the compute required for a violation shrinks.
	
	\section{Scope}
	Three components can be identified in international AI agreements: rules that define what the parties commit to; an institutional structure tasked with enforcing the rules, resolving disputes, and penalizing violations; and the technical mechanisms that concretely verify and enforce compliance, such as chip tracking, on-site inspections, and export controls. This paper addresses the third component, for agreements that rely on hardware governance to prevent violations. Within this scope, the paper analyzes technical feasibility only. Legal, commercial, and sovereignty frictions are not assessed.
	
	One approach to enforcing such rules defines a \emph{compute threshold}: a quantity of compute above which a violation becomes possible \parencite{heim2024thresholds}. All compute above this threshold must be verified to ensure it is not used to violate the agreement. This can be specified as total compute (FLOP) or as processing speed (FLOP/s), though other metrics such as total processing performance (TPP) exist \parencite{kuntamukkala2023export}. To enforce the threshold, the agreement must specify which chip types need to be tracked. In this paper, we focus on AI accelerators, while consumer GPUs and CPUs are out of scope. Whether these chips can also produce violations is left to further research. The analysis in this paper is agnostic to compute threshold height and AI accelerator chip type.
	
	\section{Taxonomy structure: three sub-problems}
	Given a compute threshold, all combinations of chips above the threshold must be monitored so that violations can be detected. We can decompose this into three sub-problems, building on \textcite{scher2025international}\footnote{The three-way split mirrors the grouping in Section 4 of \textcite{scher2025international}, which treats locating existing chips, tracking new chips, and monitoring chip use as separate concerns.}: preventing escape from the control regime, preventing uncontrolled resource acquisition, and detecting hidden capacity. Preventing escape from the control regime ensures monitored chips do not violate the agreement. The other two ensure all chips are monitored. Preventing uncontrolled resource acquisition tracks new chips, and detecting hidden capacity locates existing chips.
	
	\begin{figure}[H]
		\centering
		\includegraphics[width=\textwidth]{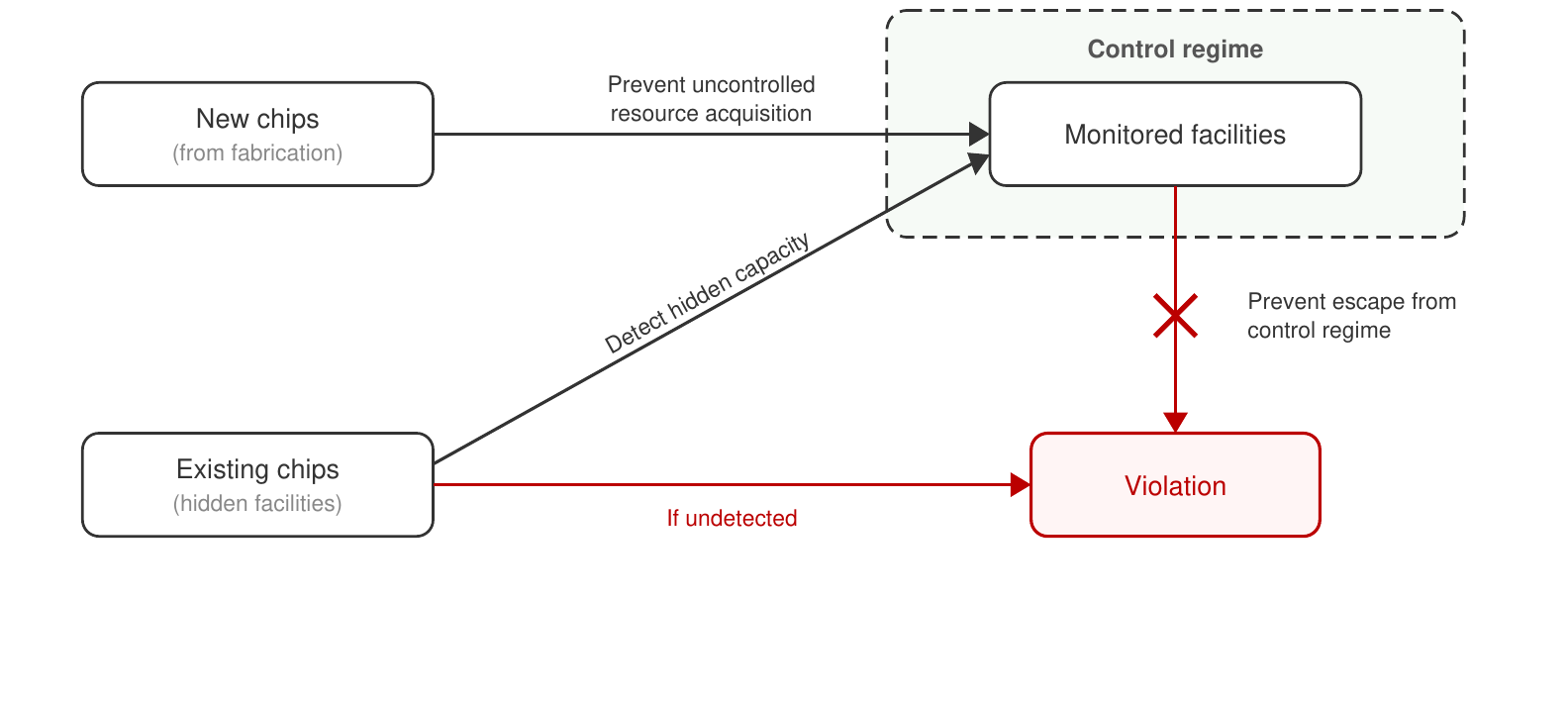}
		\caption{State diagram of the three sub-problem decomposition.}
		\label{fig:state-diagram}
	\end{figure}
	
	Figure~\ref{fig:state-diagram} illustrates this decomposition. Within the control regime, preventing escape ensures monitored facilities do not violate the agreement. Preventing uncontrolled resource acquisition tracks new chips into monitored facilities. Detecting hidden capacity locates existing chips. Most existing capacity enters the regime through declarations required at agreement entry. Detection targets what remains hidden. Once found, that capacity is brought under monitoring or dismantled. Existing chips that remain undetected can violate the agreement directly if their combined compute exceeds the threshold. Table~\ref{tab:framework} summarizes each sub-problem with its success criterion.
	
	\begin{table}[H]
		\centering
		\begin{tabular}{p{3.2cm} p{4.5cm} p{5.5cm}}
			\toprule
			\textbf{Sub-problem} & \textbf{State transition} & \textbf{Success criterion} \\
			\midrule
			Prevent escape from the control regime & Monitored facility $\to$ Violation & No monitored facility violates the agreement \\
			\midrule
			Prevent uncontrolled resource acquisition & New chips $\to$ Monitored facility & All new chips above the threshold enter monitored facilities \\
			\midrule
			Detect hidden capacity & Existing chips $\to$ Monitored facility & All existing chips above the threshold are detected and enter monitored facilities \\
			\bottomrule
		\end{tabular}
		\caption{Taxonomy of the three sub-problems.}
		\label{tab:framework}
	\end{table}
	
	\section{Existing proposals}
	\label{sec:existing-proposals}
	
	In this section, we categorize policies from eight proposals for international AI agreements into the three sub-problems. Many policies recur across proposals, and grouping them shows which sub-problems each proposal addresses. We include proposals that specify concrete enforcement or verification measures in substantial detail, and exclude those that discuss institutional design or introduce only a single policy. Some proposals specify a compute threshold (see Appendix~\ref{app:proposal-thresholds}).
	
	Each of these proposals describes multiple policies at varying levels of granularity: up to a hundred policies are listed in a single source \parencite{scher2025mechanisms}. We group policies that are sub-types of the same general approach into a single category; for example, visual and infrared satellite surveillance are listed as one. Tables~\ref{tab:proposals-escape},~\ref{tab:proposals-acquisition}, and~\ref{tab:proposals-detection} list these categories per sub-problem; a dot indicates that the proposal includes at least one policy in that category.
	
	In the next three subsections, we present an overview of identified policies per existing proposal for each sub-problem.
	
	\subsection{Policies preventing escape from the control regime}
	The policies in Table~\ref{tab:proposals-escape} aim to prevent a controlled facility from violating the agreement. We group them into five categories, each with a different role. Verifying chip activity catches prohibited workloads as they run. Auditing declared training runs catches false declarations after the fact. Hardware-level controls cap what the chip can do. Authorization pre-approves what can be run on a chip. Preparatory infrastructure supports the other four by consolidating chips into monitored facilities. Following \textcite{scher2025international}, the control regime we adopt rests on chip consolidation into monitored facilities. The sub-problem decomposition itself is agnostic about what the regime contains; any regime that detects and stops violations works.
	
	\begin{table}[H]
		\centering
		\begin{threeparttable}
			\resizebox{\textwidth}{!}{%
				\begin{tabular}{lcccccccc}
					\toprule
					& Al Ramiah\textsuperscript{1} & Scher\textsuperscript{2} & Miotti\textsuperscript{3} & Aguirre\textsuperscript{4} & Wasil\textsuperscript{5} & Harack\textsuperscript{6} & Baker\textsuperscript{7} & Scher\textsuperscript{8} \\
					\midrule
					\textbf{Verifying chip activity} & & & & & & & & \\
					\quad On-site inspectors & $\bullet$ & $\bullet$ & $\bullet$ & & $\bullet$ & $\bullet$ & $\bullet$ & $\bullet$ \\
					\quad Tamper-proof cameras & $\bullet$ & $\bullet$ & & & & $\bullet$ & $\bullet$ & $\bullet$ \\
					\quad On-chip verification mechanisms & & $\bullet$ & & & & $\bullet$ & $\bullet$ & $\bullet$ \\
					\quad Off-chip verification mechanisms & & $\bullet$ & & & & $\bullet$ & $\bullet$ & $\bullet$ \\
					\quad Challenge inspections & $\bullet$ & $\bullet$ & & & $\bullet$ & $\bullet$ & $\bullet$ & $\bullet$ \\
					\midrule
					\textbf{Auditing declared training runs} & & & & & & & & \\
					\quad Proof of training (cryptographic) & & & & $\bullet$ & & $\bullet$ & $\bullet$ & $\bullet$ \\
					\quad Rerunning declared workloads & & $\bullet$ & & & & & $\bullet$ & $\bullet$ \\
					\midrule
					\textbf{Hardware-level controls} & & & & & & & & \\
					\quad Bandwidth/latency restrictions & $\bullet$ & $\bullet$ & & $\bullet$ & & $\bullet$ & & $\bullet$ \\
					\quad FLOP-rate and precision restrictions & & $\bullet$ & & & & & & $\bullet$ \\
					\quad Kill-switches & & $\bullet$ & $\bullet$ & $\bullet$ & & & & $\bullet$ \\
					\midrule
					\textbf{Authorization} & & & & & & & & \\
					\quad Firmware-based offline licensing & $\bullet$ & & & & & $\bullet$ & & $\bullet$ \\
					\quad Pre-training authority approval & & $\bullet$ & $\bullet$ & $\bullet$ & & $\bullet$ & & \\
					\quad KYC for compute providers & $\bullet$ & & $\bullet$ & & & & & \\
					\quad Developer / user licensing & & & $\bullet$ & $\bullet$ & & & & \\
					\midrule
					\textbf{Preparatory infrastructure} & & & & & & & & \\
					\quad Chip consolidation into monitored facilities & & $\bullet$ & & & & & & \\
					\bottomrule
				\end{tabular}%
			}
			\begin{tablenotes}
				\small
				\item[1] \cite{alramiah2025globalregime}
				\item[2] \cite{scher2025international}
				\item[3] \cite{miotti2025narrow}
				\item[4] \cite{aguirre2025keepfuture}
				\item[5] \cite{wasil2024verification}
				\item[6] \cite{harack2025verification}
				\item[7] \cite{baker2025verifying}
				\item[8] \cite{scher2025mechanisms}
			\end{tablenotes}
			\caption{Proposals for sub-problem: prevent escape from the control regime.}
			\label{tab:proposals-escape}
		\end{threeparttable}
	\end{table}
	
	\subsection{Policies preventing uncontrolled resource acquisition}
	Table~\ref{tab:proposals-acquisition} lists two types of policies that aim to keep newly produced chips within the control regime. Hardware restrictions define which chips need to be controlled. Monitoring verifies that produced chips reach the control regime.
	
	\begin{table}[H]
		\centering
		\begin{threeparttable}
			\resizebox{\textwidth}{!}{%
				\begin{tabular}{lcccccccc}
					\toprule
					& Al Ramiah\textsuperscript{1} & Scher\textsuperscript{2} & Miotti\textsuperscript{3} & Aguirre\textsuperscript{4} & Wasil\textsuperscript{5} & Harack\textsuperscript{6} & Baker\textsuperscript{7} & Scher\textsuperscript{8} \\
					\midrule
					\textbf{Hardware restrictions} & & & & & & & & \\
					\quad Export controls on fab equipment & $\bullet$ & $\bullet$ & & & & & & \\
					\quad Chip production caps & $\bullet$ & $\bullet$ & & & & & & \\
					\quad Export controls on AI chips & $\bullet$ & $\bullet$ & & & & & & $\bullet$ \\
					\quad Export controls on non-AI chips & & $\bullet$ & & & & & & \\
					\quad Restricting non-party cloud access & & $\bullet$ & & & & & & \\
					\midrule
					\textbf{Monitoring} & & & & & & & & \\
					\quad Chip fab inspections & & & & & $\bullet$ & $\bullet$ & & $\bullet$ \\
					\quad Detect illegal chip trade & & $\bullet$ & & & $\bullet$ & & & \\
					\quad Chip production tracking & $\bullet$ & $\bullet$ & & $\bullet$ & & $\bullet$ & $\bullet$ & $\bullet$ \\
					\quad Chip location registry & & $\bullet$ & $\bullet$ & $\bullet$ & $\bullet$ & $\bullet$ & $\bullet$ & $\bullet$ \\
					\quad Financial intelligence & & & & & $\bullet$ & & & $\bullet$ \\
					\quad Compute usage reporting & & $\bullet$ & $\bullet$ & $\bullet$ & & & $\bullet$ & $\bullet$ \\
					\bottomrule
				\end{tabular}%
			}
			\caption{Proposals for sub-problem: prevent uncontrolled resource acquisition. Sources as in Table~\ref{tab:proposals-escape}.}
			\label{tab:proposals-acquisition}
		\end{threeparttable}
	\end{table}
	
	\subsection{Policies detecting hidden capacity}
	The policies in Table~\ref{tab:proposals-detection} aim to find facilities operating outside the control regime. They fall into two categories. Physical detection uses signals from chips and the facilities housing them. Information gathering draws on human sources to identify chip locations.
	
	\begin{table}[H]
		\centering
		\resizebox{\textwidth}{!}{%
			\begin{tabular}{lcccccccc}
				\toprule
				& Al Ramiah\textsuperscript{1} & Scher\textsuperscript{2} & Miotti\textsuperscript{3} & Aguirre\textsuperscript{4} & Wasil\textsuperscript{5} & Harack\textsuperscript{6} & Baker\textsuperscript{7} & Scher\textsuperscript{8} \\
				\midrule
				\textbf{Physical detection} & & & & & & & & \\
				\quad Satellite surveillance & & $\bullet$ & $\bullet$ & & $\bullet$ & $\bullet$ & $\bullet$ & $\bullet$ \\
				\quad Aerial surveillance & & $\bullet$ & $\bullet$ & & $\bullet$ & $\bullet$ & $\bullet$ & \\
				\quad Power consumption monitoring & & $\bullet$ & $\bullet$ & & $\bullet$ & $\bullet$ & & $\bullet$ \\
				\quad Geophysical MASINT & & & & & & & & $\bullet$ \\
				\quad Cyber/signals intelligence & & & & & & & $\bullet$ & $\bullet$ \\
				\midrule
				\textbf{Information gathering} & & & & & & & & \\
				\quad Whistleblower protections & & $\bullet$ & & & $\bullet$ & $\bullet$ & $\bullet$ & $\bullet$ \\
				\quad Human intelligence and interviews & & $\bullet$ & & & & $\bullet$ & $\bullet$ & $\bullet$ \\
				\quad Researcher employment monitoring & & $\bullet$ & & & & & & $\bullet$ \\
				\bottomrule
			\end{tabular}%
		}
		\caption{Proposals for sub-problem: detect hidden capacity. Sources as in Table~\ref{tab:proposals-escape}.}
		\label{tab:proposals-detection}
	\end{table}
	
	These three tables show substantial overlap between the policies proposed across the eight sources; most are slight variants of the same underlying approach. They do not, however, assess whether those measures remain effective as the compute threshold shrinks and more actors become capable of violating an agreement. The following section addresses that question.
	
	\section{Enforcement breaking points}
	In this section, we discuss what is needed for success for each sub-problem, and we clarify how existing policies contribute. As the facility size required to achieve a given capability shrinks, enforcing some policies becomes progressively harder. Therefore, where relevant, we also attempt to identify the facility
	size at which the solution breaks down. We find that \emph{prevent escape} and \emph{prevent uncontrolled resource acquisition} remain tractable at lower compute thresholds, while \emph{detect hidden capacity} becomes intractable.
	
	\subsection{Prevent escape from the control regime}
	\label{sec:prevent-escape}
	This sub-problem is solved when, for every facility above the compute threshold and inside the control regime, violations are prevented with high confidence. Following \textcite{scher2025international}, we treat the following regime as technically sufficient to prevent violations in the control regime. Every chip resides in a monitored facility \parencite{scher2025international}, where, initially, inspectors are given ongoing physical access to chips. The workload is continuously monitored by tamper-proof cameras \parencite{scher2025mechanisms}, off-chip power and networking measurements \parencite{baker2025verifying, scher2025mechanisms, harack2025verification}, on-chip hardware-enabled mechanisms \parencite{aarne2024securechips, petrie2024firmware, petrie2025flexible, brass2024location}, and rerunning of declared workloads \parencite{shavit2023chinchilla}. Where verification cannot provide assurance, hardware is powered off and its non-operation is continuously verified \parencite[Article VII \S3]{scher2025international}. As a last resort, the facility can be completely shut down. 
	
	We expect this regime is sufficient to solve the sub-problem. In the worst case, when monitoring and power-off leave residual uncertainty, complete shutdown remains reliable. Every chip sits in a controlled facility. This centralization lets the operator cut power to all chips within a single facility from a single point.
	
	Moreover, this sub-problem is more robust to a falling compute threshold than detecting hidden capacity. Each chip is monitored individually, so a smaller compute requirement does not by itself weaken the regime. This depends on two conditions. First, the scope is restricted to AI accelerators, which excludes legacy and general-purpose chips that lack built-in security. Second, off-chip, chip-agnostic measures might verify chips that lack on-chip security \parencite{petrie2025flexible}. Whether such methods cover every case is an open question. While feasibility of solving this sub-problem is not fundamentally threshold-dependent, enforcement costs do scale with threshold height. As it lowers, the number of actors able to breach the agreement grows, making enforcement more expensive.
	
	Several methods aim to make complete shutdown unnecessary. By verifying chip activity reliably, they let permitted workloads continue and lower ongoing monitoring costs. This raises political will, improving the prospects for adoption.
	
	These methods take several forms. Proof-of-training protocols verify what was run in a privacy-preserving way \parencite{shavit2023chinchilla}. Firmware-based offline licensing reduces ongoing inspector involvement \parencite{petrie2024firmware}. Hardware-enabled mechanisms automate enforcement at the chip level \parencite{petrie2025flexible, aarne2025flexiblehardware, kulp2024hardwareenabled, aarne2024securechips}. Workload classification and the rerunning of declared workloads provide further automated checks \parencite{scher2025mechanisms}. Table~\ref{tab:proposals-escape} catalogs these measures.
	
	\subsection{Prevent uncontrolled resource acquisition}
	\label{sec:acquisition}
	This sub-problem is solved when all chips capable of contributing to a violation enter the control regime, where the solution in Section~\ref{sec:prevent-escape} applies. Solving it also makes \emph{detect hidden capacity} (Section~\ref{sec:detect}) easier, since chips no longer reach hidden facilities.
	
	For this sub-problem, we likewise treat the regime described below as technically sufficient. It rests on chip tracking grounded in the concentrated AI chip supply chain. TSMC dominates the advanced packaging (CoWoS) used in frontier AI accelerators, with no current alternative at comparable scale or yield. Only three firms (SK Hynix, Samsung, Micron) produce high-bandwidth memory, ASML alone makes EUV lithography machines, and TSMC fabricates almost all advanced logic \parencite{sastry2024computing, epoch_chip_components}. This concentration makes production monitoring feasible.
	
	Three elements of the regime are relevant. First, chip tracking: every new AI chip in scope is tracked from fabrication through testing, packaging, and installation in a controlled facility \parencite{shavit2023chinchilla, brass2024location, sastry2024computing}. Second, transfer controls: chip transfers outside the regime are presumed denied \parencite{allen2022choking}. Third, fab controls: production at any fab that cannot be tracked is halted \parencite{scher2025international}, and transfers of chip-manufacturing equipment outside the regime are presumed denied \parencite{allen2022choking}.
	
	We expect, as of today, this regime is sufficient to solve the sub-problem. In the worst case, production of new chips can be halted. Only a handful of fabs, packaging facilities, HBM producers, and EUV suppliers exist worldwide, which makes halting feasible. The same concentration enables ongoing monitoring of every chip, making the worst case unlikely.
	
	As the threshold lowers, the first two elements (chip tracking and transfer controls) remain effective. Each chip is tracked from fabrication and cannot be legitimately transferred outside the regime. Acquiring chips outside the agreement, for example during transport, becomes easier when required amounts are lower, yet the approach does not fundamentally break down at any quantifiable cluster size. The sub-problem's solution is therefore in principle threshold-agnostic.
	
	In contrast, the third element (fab controls) weakens as the threshold lowers. Today, covert fabs cannot produce frontier AI chips, because the required complexity and size make their construction infeasible. This is likely to hold in the coming years, so the gap is not immediately pressing.
	
	As the threshold falls further, however, the required complexity and size drop, and covert fab construction becomes feasible. Even then, such fabs remain larger and easier to detect than the facilities Section~\ref{sec:detect} must find at the same threshold, so \emph{detect hidden capacity} becomes the binding constraint before fab controls do. Tracking chip-manufacturing equipment alongside chips would strengthen fab controls as the threshold lowers further.
	
	\subsection{Detect hidden capacity}
	\label{sec:detect}
	
	\emph{Detect hidden capacity} succeeds when no facility that exceeds
	the threshold is unknown. If \emph{prevent uncontrolled resource acquisition}
	(Section~\ref{sec:acquisition}) is solved, detection is needed only at
	agreement entry because no new chips can enter unmonitored channels
	afterwards.
	
	Unlike the other two sub-problems, detecting hidden capacity becomes intractable as the threshold falls. Smaller facilities are harder to distinguish from ordinary buildings, and the number of candidate buildings grows. Detection depends on both the physical limits of each method and whether the facility is deliberately hidden. We first analyze what detection methods achieve on their own, then consider how an adversary can defeat them.
	
	\subsubsection{Detection method breaking points}
	Detection methods fall into two approaches. First, information gathering draws on public documents, sales records, human intelligence, and whistleblowers to locate facilities. Second, physical detection identifies facilities directly through their distinguishing signatures: physically distinct buildings and large power consumption. 
	
	Epoch AI has combined both approaches to catalog frontier AI data centers, focusing on facilities large and visible enough to verify through public sources. It identified facilities housing 15\% of the 20.2 million frontier AI chips worldwide \parencite{epoch_data_centers, epoch_chip_owners}. Publicly available sources and prior research identified candidates. Permits, company documents, and satellite imagery confirmed facility details \parencite{epoch_data_centers_methodology}. This effort targeted only large facilities. Information gathering and physical detection can each detect more when applied comprehensively.
	
	Information gathering can trace most chips to their owners, though it might not directly locate a facility. Epoch AI traced 15.7 million of the 20.2 million chips to their owning company \parencite{epoch_chip_owners}. The remaining 4.5 million split into 1.12 million attributed to China and 3.36 million to other owners, a mix of smaller known companies and unknowns. Hundreds of thousands to millions of chips have no known owner, leaving room for hidden facilities. Other information-gathering methods such as building permits, human intelligence, and whistleblowers might bridge this gap. However, available information decreases as facilities shrink. Smaller facilities employ fewer people and have a lower public profile, leaving fewer traces.
	
	Physical detection attempts to close this gap. Satellite surveillance reliably detects facilities above 10~MW ($\sim$10,000~H100-equivalents \parencite{nvidia_dgx_h100_datasheet}) and might fail below this range \parencite{clymer_classifier_repo}. The key feature that separates data centers from other buildings is their cooling infrastructure \parencite{epoch_data_centers_methodology}. A 10,000~H100-equivalent facility carries large water-cooled chiller plants, unusual for a building of its size. At 1,000~H100-equivalents air-cooled chillers are more common \parencite{vertiv_smartmod_datasheet, schneider_ecostruxure_designs} and harder to distinguish. At 100~H100-equivalents the facility fits inside a 12-meter container with no noticeable external cooling \parencite{vertiv_smartmod_datasheet}, removing the distinguishing feature entirely. Aerial surveillance offers closer detail, yet faces the same problem. Without a distinguishing feature, detection becomes infeasible.
	
	Power consumption monitoring degrades on the same path. At 10,000~H100-equivalents the facility draws 10~MW continuously. \textcite{pilz2023compute} estimate 335 to 1,325 data centers globally above 10~MW (70\% confidence interval). Their report sets no 1,000-equivalent threshold, but estimates 10,000 to 30,000 facilities above 0.1~MW (100~H100-equivalents). The 1,000-equivalent count must lie between this figure and the 10~MW count above, so between roughly 1,325 and 30,000 globally. These figures cover only labeled data centers. A 100~kW load sits within the existing population of mid-size commercial customers, which number in the millions \parencite{eia2018cbecs_b6}.
	
	Signals intelligence and geophysical MASINT likely face the same limitation. We do not analyze either method in detail. Appendix~\ref{app:detection-detail} discusses both alongside the other physical detection methods.
	
	In conclusion, facilities at 10,000~H100-equivalents are still mostly detectable, primarily through satellite surveillance. Combining methods that are independent of each other raises the detection probability further. Below this scale, distinguishability drops sharply and detection becomes much harder. 
	
	\subsubsection{Active concealment}
	These estimates assume the facility is not deliberately hidden. Active concealment makes detection harder. Even at 10,000 or 100,000~H100-equivalents, hidden facilities cannot be ruled out with full confidence. Nonetheless, concealing a 100,000~H100-equivalent facility is difficult.
	
	A rough upper bound follows from production figures. As noted earlier, 15.7 million of the 20.2 million chips produced can be traced to the owning company. Much of the remaining 4.5 million can likely also be traced, leaving on the order of a million chips unaccounted for. This yields an approximate bound on the size of the largest hidden cluster.
	
	Whistleblower programs and other incentives to surface hidden data centers can complement detection in this scenario. The exact lower bound depends on what intelligence agencies achieve by combining all detection methods, which is hard to estimate from public information.
	
	\begin{table}[H]
		\centering
		\resizebox{\textwidth}{!}{%
			\begin{threeparttable}
				\begin{tabular}{@{}l cccc l@{}}
					\toprule
					& \multicolumn{4}{c}{\textbf{Facility size (H100-equivalents)}} & \textbf{Deployment} \\
					\cmidrule(lr){2-5}
					& \textit{100,000} & \textit{10,000} & \textit{1,000} & \textit{100} & \textbf{requirement} \\
					\midrule
					\textbf{Physical detection} & & & & & \\
					\quad Satellite surveillance & & & & & Unilateral \\
					\qquad No concealment & \emptycirc & \emptycirc & \halfcirc & \fullcirc & \\
					\qquad Active concealment & \halfcirc & \fullcirc & \fullcirc & \fullcirc & \\
					\quad Aerial surveillance & & & & & Airspace access \\
					\qquad No concealment & \emptycirc & \emptycirc & \halfcirc & \fullcirc & \\
					\qquad Active concealment & \halfcirc & \fullcirc & \fullcirc & \fullcirc & \\
					\quad Power consumption monitoring & & & & & Utility cooperation \\
					\qquad No concealment & \emptycirc & \halfcirc & \halfcirc & \fullcirc & \\
					\qquad Active concealment & \halfcirc & \fullcirc & \fullcirc & \fullcirc & \\
					\midrule
					\textbf{Information gathering} & & & & & Various \\
					\multicolumn{6}{l}{\small\textit{Whistleblower programs, human intelligence, researcher employment monitoring,}} \\
					\multicolumn{6}{l}{\small\textit{publicly available information, building permits, sales records, company documents.}} \\
					\quad No concealment & \quartercirc & \quartercirc & \halfcirc & \halfcirc & \\
					\quad Active concealment & \halfcirc & \halfcirc & \threequartercirc & \threequartercirc & \\
					\bottomrule
				\end{tabular}
				\begin{tablenotes}
					\small
					\item \fullcirc\ = measure fails \quad \threequartercirc\ = measure mostly fails \quad \halfcirc\ = measure partially fails \quad \quartercirc\ = measure mostly works \quad \emptycirc\ = measure works
					\item The ``No concealment'' row assumes no deliberate concealment effort. The ``Active concealment'' row reflects an adversary actively concealing the facility.
				\end{tablenotes}
			\end{threeparttable}%
		}
		\caption{Enforcement breaking point analysis for sub-problem: detect hidden capacity.}
		\label{tab:detection}
	\end{table}
	
	\section{Conclusion}
	Controlling all hardware above a compute threshold is a promising way to enforce international AI agreements. We decomposed this problem into three sub-problems, categorized existing measures under them, specified what it takes to solve each, and analyzed whether they remain tractable as the threshold lowers.
	
	\emph{Detect hidden capacity} is the sub-problem most likely to become intractable as the threshold lowers. The other two sub-problems verify chips whose existence is already known. This one must find facilities that are not. The signals that could reveal them, both physical signatures and other locating information, weaken as facility size shrinks. Without active concealment, facilities remain mostly detectable down to roughly 10,000~H100-equivalents, and detection degrades sharply below that scale. With active concealment, hidden facilities of 10,000 to 100,000~H100-equivalents cannot be ruled out.
	
	The other two sub-problems are more robust to a falling threshold. \emph{Prevent uncontrolled resource acquisition} relies on a concentrated supply chain, which is unlikely to disperse in the coming years. \emph{Prevent escape} monitors each chip individually, so a smaller compute threshold does not by itself weaken the regime. Both still have unsolved components, yet their difficulty grows less steeply than that of \emph{detect hidden capacity} as the threshold lowers.
	
	\emph{Detect hidden capacity} is thus the sub-problem that determines whether verification succeeds. If hidden facilities cannot be found, the regime monitors only the compute that actors choose to show. However, this does not imply that research should concentrate on detection, since traction there may be harder to gain than on the other two sub-problems. All three therefore deserve continued attention.
	
	\section{Limitations}
	This analysis has several important limitations, each marking a boundary of the current work where a fuller treatment would sharpen or extend the conclusions. We detail them below.
	
	\begin{description}[leftmargin=0pt, style=unboxed, font=\normalfont\bfseries, topsep=0pt, parsep=0pt, itemsep=0.55\baselineskip]
		\item[Deliberately broad.] This paper serves as an overview of what verification and enforcement consist of and what a solution requires. Details of how each sub-problem is solved are left to future work and the cited proposals.
		\item[Legal and commercial barriers not analyzed.] The analysis treats verification as a technical-feasibility question. Legal, commercial, and sovereignty frictions may bind earlier than technical limits.
		\item[Mutual verification not modeled.] The paper assumes that solutions to the three sub-problems are successfully implemented. It does not cover, for example, how the US and China would verify that the other party's domestic enforcement is executed correctly.
		\item[Response to detected violations.] The paper treats detection of a violation as the endpoint of the analysis. What response follows detection, and what is required for that response to deter violations, is not specified.
		\item[Single analysis behind the 10,000~H100-equivalent number.] The number rests on a single open-research analysis. One such study suffices to demonstrate that detection is possible at that scale. However, it does not establish a lower bound on the smallest detectable facility. Smaller facilities may be detectable with more effort, and intelligence agencies may have capabilities not reflected in open research.
		\item[Active concealment under-investigated.] The range of what is possible under active concealment carries large uncertainty.
	\end{description}

	\section{Future work}
	
	Our analysis shows \emph{detect hidden capacity} is the sub-problem most likely to break when the threshold lowers. Better detection methods would make it more tractable, alongside resolving the open components of the other two sub-problems. How urgent this work is depends on how fast the threshold falls. Two trends push it down. First, algorithmic progress lowers the compute required for dangerous capabilities, so the threshold must be set lower over time to keep covering them. Second, hardware progress and low-bandwidth training make a given threshold harder to maintain. The same quantity of compute then fits in smaller facilities and runs on consumer GPUs and distributed hardware. The pace of both trends determines how long current enforcement measures remain viable.
	
	\begin{description}[leftmargin=0pt, style=unboxed, font=\normalfont\bfseries, topsep=0pt, parsep=0pt, itemsep=0.55\baselineskip]
		\item[Provide more in-depth solutions for prevent escape and prevent uncontrolled resource acquisition.] Solutions to these two sub-problems have been proposed but are not yet finalized. Implementation details and cost reduction remain for future work.
		\item[New and combined detection approaches.] Existing detection measures struggle at small facility sizes. Whether this reflects the limits of current proposals or a fundamental constraint on detectability is an open question. Future work should investigate whether novel or combined detection methods can extend the detection window to smaller cluster sizes, or alternatively establish lower bounds on the facility size at which detection becomes intractable regardless of method.
		\item[Slowing down the lowering of the threshold.] This paper evaluates enforcement at several threshold levels but treats each as a snapshot. It does not investigate policies to slow down this descent. Analyzing restrictions such as research bans would quantify how much they can slow this descent.
		\item[Institutional design.] This paper addresses the technical feasibility of enforcement. It does not address who operates the control regime, how it is governed, what triggers a response to a detected violation, or how disputes between signatories are resolved. These institutional questions determine whether technically feasible enforcement can be realized in practice, and constitute a separate line of research.
	\end{description}
	
	\newpage
	\printbibliography[title=Literature]
	
	\newpage
	\appendix
	
	\section{Proposal thresholds}
	\label{app:proposal-thresholds}
	Table~\ref{tab:metrics} lists the compute thresholds specified by the eight proposals categorized in Section~\ref{sec:existing-proposals}. Section~\ref{sec:existing-proposals} maps which policies each proposal includes. This appendix completes that mapping with the thresholds those policies must enforce.
	
	Four proposals specify thresholds, while the other four focus on verification mechanisms rather than complete agreements and leave the threshold open. The table distinguishes total training compute (FLOP), post-training compute, and processing speed (FLOP/s). The strictness distinguishes strict thresholds, above which the activity is prohibited, from regulated thresholds, above which it requires licensing or monitoring. 
	
	\begin{table}[H]
	\centering
	\resizebox{\textwidth}{!}{%
		\begin{threeparttable}
			\begin{tabular}{lcccccccc}
				\toprule
				\textbf{Threshold type} & Al Ramiah\textsuperscript{1} & Scher\textsuperscript{2} & Miotti\textsuperscript{3} & Aguirre\textsuperscript{4} & Wasil\textsuperscript{5} & Harack\textsuperscript{6} & Baker\textsuperscript{7} & Scher\textsuperscript{8} \\
				\midrule
				Strict FLOP & $10^{25}$ & $10^{24}$ & $10^{27}$ & $10^{27}$ & \na & \na & \na & \na \\
				Strict Post-train.\ FLOP & \na & $10^{23}$ & \na & \na & \na & \na & \na & \na \\
				Strict FLOP/s & \na & \na & $10^{19}$ & $10^{20}$ & \na & \na & \na & \na \\
				Regulated\textsuperscript{b} FLOP & \na & $10^{22}$ & $10^{25}$ & $10^{25}$ & \na & \na & \na & \na \\
				Regulated\textsuperscript{b} Post-train.\ FLOP & \na & \na & Yes\textsuperscript{a} & \na & \na & \na & \na & \na \\
				Regulated\textsuperscript{b} FLOP/s & \na & 16 H100 & $10^{17}$ & $10^{18}$ & \na & \na & \na & \na \\
				\bottomrule
			\end{tabular}
			\begin{tablenotes}
				\small
				\item[1] \cite{alramiah2025globalregime}
				\item[2] \cite{scher2025international}
				\item[3] \cite{miotti2025narrow}
				\item[4] \cite{aguirre2025keepfuture}
				\item[5] \cite{wasil2024verification}
				\item[6] \cite{harack2025verification}
				\item[7] \cite{baker2025verifying}
				\item[8] \cite{scher2025mechanisms}
				\item[a] Application License: triggered by capability-increasing modifications rather than a specific FLOP threshold.
				\item[b] ``Regulated'' denotes a licensing regime for Miotti et al.\ and Aguirre, and a monitoring regime for Scher et al.
			\end{tablenotes}
		\end{threeparttable}%
	}
	\caption{Proposal thresholds.}
	\label{tab:metrics}
\end{table}
	
	\section{Further analysis of detection measures}
	\label{app:detection-detail}
	
	Section~\ref{sec:detect} reports where each detection measure breaks down. This appendix provides the underlying analysis for satellite surveillance, aerial surveillance, and power consumption monitoring. Each of these measures depends on a signature that distinguishes a compute facility from ordinary buildings. The analysis traces how that signature weakens as facility size shrinks from 10,000 to 100~H100-equivalents, and notes the deployment requirements that constrain each measure in adversarial settings. These results underlie the breaking points in Table~\ref{tab:detection}. Cyber/signals intelligence and geophysical MASINT are treated qualitatively. Both rely on signatures that concealment can suppress, and we do not estimate breaking points for them.
	
	\paragraph{Satellite surveillance.}
	A 10,000~H100-equivalent facility has a heat load of approximately 10~MW \parencite{nvidia_dgx_h100_datasheet} and rejects it through cooling-tower arrays or large water-cooled chiller plants. At this scale the cooling towers take the form of tall ground-mounted arrays, distinct from the smaller rooftop packages found on some commercial buildings. This equipment family is only economic above 5 to 10~MW \parencite{semianalysis2024cooling, ced2018hvac}. The facility itself fits in roughly 2,000~m\textsuperscript{2} \parencite{clymer_classifier_repo}, while cooling towers concentrate in much larger buildings: 43\% of US commercial buildings above 18,500~m\textsuperscript{2} (roughly 200,000~ft\textsuperscript{2}) have a cooling tower, against 3\% of buildings at or below that size \parencite{eia2018cbecs_cooling}. A cooling-tower array on a 2,000~m\textsuperscript{2} footprint is therefore anomalous.
	
	At 1,000~H100-equivalents the cooling kit drops below the cooling-tower threshold. Vendor reference designs at 1~MW use air-cooled chillers with no cooling towers (Vertiv MegaMod, Schneider EcoStruxure Pod) \parencite{vertiv_megamod_coolchip, schneider_ecostruxure_designs}. Air-cooled chillers are long horizontal boxes mounted on a pad or roof, each rated at a few hundred tons of capacity. The same equipment is found on hospitals, hotels, supermarkets, and mid-size offices: central chillers appear on 3\% of US commercial buildings and 19\% of commercial floorspace \parencite{eia2018cbecs_b6}. A chiller plant on a small or mid-size commercial shell is unremarkable.
	
	At 100~H100-equivalents the facility fits inside a 12-meter container. The Vertiv SmartMod ships at this scale with two to six in-row direct-expansion units rejecting heat through outdoor condensers \parencite{vertiv_smartmod_datasheet}, with the same form factor as commercial DX units of equivalent capacity. Packaged air-conditioning units of this kind appear on 43\% of US commercial buildings and 58\% of commercial floorspace \parencite{eia2018cbecs_b6}. There is no externally distinctive signature.
	
	Clymer's empirical results are consistent with this transition. All four false negatives in his classifier had power capacity below 10~MW (1.2, 0.5, 3.2, and 0.5~MW), with the smallest annotated as a small university cluster \parencite{clymer_classifier_repo}. Above 10~MW the classifier achieved a 100\% true positive rate; below it, the same classifier missed every facility in the available sample.
	
	Satellite surveillance requires no cooperation from the target state, making it deployable unilaterally. It is most useful as a first filter for facilities at 10,000~H100-equivalents and above, with corroboration from other measures.
	
	\paragraph{Aerial surveillance.}
	Aerial platforms such as drones and manned aircraft offer
	centimeter-scale resolution and oblique viewing angles, which
	resolve some of the ambiguity that limits satellite imagery at
	smaller facility sizes. However, like satellites they see only the
	exterior of the building, with greater precision. The analysis for
	satellite surveillance therefore applies similarly.
	
	A second difference is the deployment constraint. Aerial
	surveillance requires airspace access over the target area. A
	state actor can deny overflight, which specifically weakens aerial
	surveillance against the actor type most likely to operate large
	concealed facilities. Where satellite surveillance can be deployed
	unilaterally, aerial surveillance depends on either cooperation or
	covert operation, limiting its use in adversarial scenarios.
	
	\paragraph{Power consumption monitoring.}
	Power monitoring analyzes utility electricity records for consumption patterns characteristic of compute facilities: high, continuous, flat load with little diurnal variation. Most commercial buildings have substantial diurnal variation; data centers, cryptocurrency mining sites, refrigeration-heavy commercial buildings, telecom central offices, and continuous-process industrial customers do not.
	
	At 10,000~H100-equivalents the facility draws approximately 10~MW continuously \parencite{nvidia_dgx_h100_datasheet}. \textcite{pilz2023compute} estimate 335 to 1,325 data centers globally above 10~MW (70\% confidence interval), and flag this as likely an undercount given missing enterprise data centers and Chinese facilities. The reference class is broader still: the EIA has identified 137 US cryptocurrency mining sites with combined estimated capacity of 10,275~MW and individual sites up to 500~MW \parencite{eia2024crypto}, alongside continuous-load industrial customers (smelters, fabs, paper mills, chemical plants) and large hospital and university campuses. A 10~MW continuous load is visible at the utility level but blends with this reference class; load magnitude alone does not identify a facility as AI-specific.
	
	At 1,000~H100-equivalents the facility draws approximately 1~MW continuously. Pilz and Heim do not report a count at this threshold. Their figures of 335 to 1,325 data centers above 10~MW and 10,000 to 30,000 above 0.1~MW bracket it between roughly 1,325 and 30,000 globally. The reference class extends beyond labeled data centers to mid-size commercial customers with continuous load: large hospitals, hotels, supermarkets with extensive refrigeration, cold-storage warehouses, and telecom central offices. Load magnitude at this scale does not distinguish the facility from this reference class.
	
	At 100~H100-equivalents the facility draws approximately 100~kW continuously. Pilz and Heim estimate 10,000 to 30,000 data centers globally above 0.1~MW, but this count covers only labeled data centers. A 100~kW continuous load fits within the existing population of mid-size commercial customers with refrigeration, telecom, or 24/7 operations, of which the United States has millions \parencite{eia2018cbecs_b6}. There is no externally distinguishing power signature at this scale.
	
	Active concealment is relatively straightforward at any scale: an actor can generate power off-grid, which BYOG-style policies may normalize \parencite{paullin2026byog}. Power monitoring also requires utility cooperation, which limits its deployment in adversarial scenarios. Its added value is as a complement to satellite and aerial surveillance: a facility that defeats visual detection by resembling a normal building may still show anomalous power draw, and vice versa. The two channels are independent, so combining them narrows the candidate set more than either alone.
	
	\paragraph{Cyber/signals intelligence.}
	Signals intelligence monitors network traffic for patterns characteristic of large training runs. The distinctive traffic of a training cluster, such as gradient synchronization and parameter updates between nodes, occurs on the internal network and is not visible externally. External traffic (dataset downloads, library updates, checkpoint uploads) is modest and intermittent and may provide a weak signal, yet concealment effort can eliminate it: air-gapping the training network or routing external traffic through standard channels removes the signature.
	
	\paragraph{Geophysical MASINT.}
	Geophysical measurement and signature intelligence detects physical phenomena such as seismic vibrations from cooling compressors and fans, acoustic noise, and electromagnetic emissions from computing equipment. Unlike satellite surveillance, MASINT requires sensors deployed near the suspected facility. It is a confirmation tool, not a search tool \parencite{odni2022masint}.
	
	The method is especially helpful for detecting underground facilities where visual surveillance fails. Seismic monitoring can reveal subsurface activity that no optical sensor can reach. Even in this niche, vibration dampening could defeat the signal.
\end{document}